\documentstyle[12pt]{article}    
\begin{document}           
\rm
\baselineskip=0.33333in
\begin{quote} \raggedleft TAUP 2186-94
\end{quote}
\vglue 0.5in
\begin{center}{\bf Charges, Monopoles and \\
Duality Relations}
\end{center}
\begin{center}E. Comay
\end{center}

\begin{center}
School of Physics and Astronomy \\
Raymond and Beverly Sackler Faculty \\
of Exact Sciences \\
Tel Aviv University \\
Tel Aviv 69978 \\
Israel
\end{center}
\vglue 0.8in
\noindent

\vglue 0.8in
\noindent
Abstract:

A charge-monopole theory is derived from simple and self-evident
postulates. Charges and monopoles take an analogous
theoretical structure. It is proved that charges interact with
free waves emitted from monopoles but not with the corresponding
velocity fields. Analogous relations hold for monopole equations
of motion. The system's equations of motion can be derived
from a regular Lagrangian function.

\newpage
\noindent
{\bf 1. Introduction}
\vglue 0.33333in
    The problem of a theoretical formulation of
the behavior of a charge-monopole system is known for a
long time. Many discussions on the subject can be found
in the literature[1,2]. One unsettled point of many charge-monopole
theories is the lack of a regular classical Lagrangian
function from which the theory can be derived[3,4]. This
shortcoming entails difficulties in a construction
of a quantum mechanical theory of the system[5,6].

   An attempt to overcome this problem has been carried
out several years ago[4]. This approach relies on the
experimental fact which says that monopoles have not been
seen in laboratories. Hence, their equations of motion
have not been confirmed in experiment. It follows that
one is not obliged to construct the theory from a specific
form of the equations of motion. Thus, the approach of [4]
is based on the following postulates:
\begin{itemize}
\item[{A.}] A system of monopoles and fields is dual to a
system of charges and fields.
\item[{B.}] The equations of motion of a system of charges,
monopoles and fields can be derived from a regular Lagrangian
function.
\end{itemize}
In addition to these postulates, several other
self-evident ones are used.
It is shown in [4] that one can construct a self-consistent
theory on this basis. As postulated, this theory is derived
from a regular Lagrangian function.

    The purpose of the present work is to investigate the
possibility of constructing the same theory without relying on
postulate B. Although the assumption of the existence of
a regular Lagrangian function is appealing in its own, it
is clear that the results are stronger if they are obtained
on the basis of fewer postulates. In particular, the existence
of a regular Lagrangian function is {\em deduced} from the
set of postulates used here.

    In the present work, expressions are written in units
where the speed of light $c$ equals unity. Greek indices
range from 0 to 3. The metric is diagonal and its entries
are (1,-1,-1,-1). The symbol $_,\nu $ denotes the partial
differentiation with respect to $x^\nu $. $\tau $
denotes the invariant time.

   The paper is organized as follows. Theories of two
different kinds of charges are presented in the second
sections. A theory of a combined system of charges and
monopoles is derived in the third section. Consequences
of the results are discussed in the fourth section.
The last section contains concluding remarks.

\vglue 0.6666666in
\noindent
{\bf 2. Two Kinds of Charges}
\vglue 0.33333in

   Consider a system of one kind of charges and of their
associated fields which satisfy the equations of motion
of classical electrodynamics. Here charges are
the well known electric charges. The Lorentz force exerted
on charges is
\begin{equation}
\frac {dP^\mu _{(e)}} {d\tau } = F_{(e)}^{\mu \nu }J_{(e)\nu }.
\label{eq:LORE}
\end{equation}
The subscript $(e)$ denotes that the quantities belong to
a system of charges. The entries of the fields tensor of
$(\!\!~\ref{eq:LORE})$ are
\begin{equation}
F_{(e)}^{\mu \nu } = \left(
\begin{array}{cccc}
0   & -E_x & -E_y & -E_z \\
E_x &  0   & -B_z &  B_y \\
E_y &  B_z &   0  & -B_x \\
E_z & -B_y &  B_x &  0
\end{array}
\right).
\label{eq:FMUNU}
\end{equation}

   As is well known, physics is a science
which relies on experiment. Hence, physical quantities
should be related to measurements. The Lorentz force
$(\!\!~\ref{eq:LORE})$ can be used for measuring the
fields. This objective can be achieved by means of a set
of measurements that yield unique values of the six independent
entries of the antisymmetric fields tensor $F_{(e)}^{\mu \nu }$.
Thus, the fields are {\em defined} by means of
$(\!\!~\ref{eq:LORE})$[7].

    The fields satisfy Maxwell equations
\begin{equation}
F^{\mu \nu}_{(e)\;,\nu} = -4\pi J^\mu _{(e)}
\label{eq:MAXEIH}
\end{equation}
\begin{equation}
F^{*\mu \nu}_{(e)\;,\nu} = 0.
\label{eq:MAXEH}
\end{equation}
Here
\begin{equation}
F_{(e)}^{*\mu \nu } = \frac {1}{2}\varepsilon ^{\mu \nu \alpha \beta }
F_{(e)\alpha \beta }= \left(
\begin{array}{cccc}
0   & -B_x & -B_y & -B_z \\
B_x &  0   &  E_z & -E_y \\
B_y & -E_z &   0  &  E_x \\
B_z &  E_y & -E_x &  0
\end{array}
\right)
\label{eq:FSMUNU}
\end{equation}
and $\varepsilon ^{\mu \nu \alpha \beta }$ is the completely
antisymmetric unit tensor of the fourth rank. The fields
can be derived from a regular 4-potential[8]
(Point singularities associated with an elementary classical
point charge are not discussed in the present work)
\begin{equation}
F_{(e)\mu \nu } = A_{(e)\nu ,\mu } - A_{(e)\mu ,\nu }.
\label{eq:POTE}
\end{equation}

    An important quantity is the fields energy-momentum
tensor. This quantity represents the
density and the current of energy
and momentum throughout space-time. Its form is
(see [7], p.605 or [8], p.81)
\begin{equation}
T_{(e)}^{\mu \nu } =
\frac {1}{4\pi }(F_{(e)}^{\mu \alpha }
F_{(e)}^{\beta \nu }g_{\alpha \beta }
+\frac {1}{4}F_{(e)}^{\alpha \beta }F_{(e)\alpha \beta }g^{\mu \nu }).
\label{eq:EMT}
\end{equation}

    Let us consider another system of charges whose type is
not known yet. The second system is isolated from the electromagnetic
system described above and is denoted by the
subscript $(m)$. Interactions within the $(m)$ system are
formally the same as those of the electromagnetic system $(e)$.
Thus, the foregoing laws hold for the new system and the sole
change is the replacement of the subscript $(e)$ by $(m)$.

     Like the fields of the electromagnetic system $(e)$, the
fields of the $(m)$ system are defined by the force exerted
on a charge. This quantity is written in an ordinary tensorial
form which is analogous to $(\!\!~\ref{eq:LORE})$
\begin{equation}
\frac {dP^\mu _{(m)}} {d\tau } = F_{(m)}^{\mu \nu }J_{(m)\nu }.
\label{eq:LORM}
\end{equation}
The fields equations of motion take the form of Maxwell
equations
\begin{equation}
F^{\mu \nu}_{(m)\;,\nu} = -4\pi J^\mu _{(m)}
\label{eq:MAXMIH}
\end{equation}
\begin{equation}
F^{*\mu \nu}_{(m)\;,\nu} = 0.
\label{eq:MAXMH}
\end{equation}
These fields can be derived from a regular potential which
is a 4-vector
\begin{equation}
F_{(m)\mu \nu } = A_{(m)\nu ,\mu } - A_{(m)\mu ,\nu }.
\label{eq:POTM}
\end{equation}
Other quantities of the $(m)$ system, like
the energy-momentum tensor, are also analogous to
the corresponding ones of the $(e)$ system.

    So far nothing has been said on the interrelations between
the $(e)$ system and the $(m)$ one. Three possible relations
are as follows.
\begin{itemize}
\item Charges of the $(m)$ system are electric charges.
\item Charges and fields of the $(m)$ system do not interact
with the electromagnetic system $(e)$.
\item Charges of the $(m)$ system are magnetic monopoles.
\end{itemize}

    In the first case, there is nothing to add, because
problems of electrodynamics of a
pure system of charges is outside
the scope of the present work. In the second case, the $(m)$
system cannot be seen in electromagnetic processes and
it behaves as a dark matter. The rest of this work is devoted
to interrelations between electric charges and magnetic
monopoles.

\vglue 0.6666666in
\noindent
{\bf 3. Interactions of a Charge-Monopole System}
\vglue 0.33333in

Consider a system $\{e\}$
of charges without monopoles and
a system $\{m\}$ of monopoles without charges.
One can imagine that these systems are
very far from each other. At the limit of infinite distance
the two systems do not interact and the theory of the combined
system is just the two separate theories described in
the previous section. As $\{e\}$ and $\{m\}$ move towards
each other the interaction between these systems increases.
Thus, the theory of the combined system contains the theories
of $\{e\}$ and of $\{m\}$ as two sub-theories obtained in
the limit where the two subsystems are infinitely far apart.

    It can be shown[1] that duality transformations cast
electrodynamics of charges into electrodynamics of monopoles.
Thus, in the notation used here, these transformations take
the form
\begin{equation}
\mbox{\boldmath $E$}_{(e)}\rightarrow \mbox{\boldmath $B$}_{(m)},
\label{eq:DUALITYE}
\end{equation}
\begin{equation}
\mbox{\boldmath $B$}_{(e)}\rightarrow \mbox{\boldmath $-E$}_{(m)},
\label{eq:DUALITYB}
\end{equation}
and
\begin{equation}
e \rightarrow g
\label{eq:EG}
\end{equation}
where $g$ denotes the strength of magnetic charges. In the
previous section, notation of the entries of the fields tensor
of the $\{m\}$ system is not specified. Thus, its
entries are denoted here like those of
the dual tensor $(\!\!~\ref{eq:FSMUNU})$ of the
$\{e\}$ system
\begin{equation}
F_{(m)}^{\mu \nu } = \left(
\begin{array}{cccc}
0   & -B_x & -B_y & -B_z \\
B_x &  0   &  E_z & -E_y \\
B_y & -E_z &   0  &  E_x \\
B_z &  E_y & -E_x &  0
\end{array}
\right).
\label{eq:FMMUNU}
\end{equation}
In this way the {\em notation} of charges'
and monopoles' fields, which are defined by
$(\!\!~\ref{eq:LORE})$ and $(\!\!~\ref{eq:LORM})$,
respectively, take a
related form as seen in the ordinary tensors
$(\!\!~\ref{eq:FMUNU})$ and
$(\!\!~\ref{eq:FMMUNU})$, respectively.

    In the rest of this section, it is found how the $\{e\}$
and $\{m\}$ systems can interact {\em without destroying their
intrinsic properties as formulated in the previous section}.
To this end, the problem is divided into two parts. The
linearity of classical electrodynamics enables the split of
the fields into two subsets. One of these subsets is made
of the velocity fields of charges and the other subset is
made of the acceleration fields of charges. An analogous
division takes place in the case of monopoles' fields.

    Velocity fields and acceleration fields are seen in the
Lienard-Wiechert expression for retarded field of a
charge $q$ (see [7], p. 657 or [8], p.162)
\begin{equation}
\mbox{\boldmath $E$}= q[\frac{1-v^{2}}
{(R-\mbox{\boldmath $R\cdot v$})^{3}}
(\mbox{\boldmath $R$}-R\mbox{\boldmath $v$}) +
\frac{1}
{(R-\mbox{\boldmath $R\cdot v$})^{3}}
\mbox{\boldmath $R\times $}\{(
\mbox{\boldmath $R$}-R\mbox{\boldmath $v$})
\mbox{\boldmath $\times a$}\}]
\label{eq:LWE}
\end{equation}
\begin{equation}
\mbox{\boldmath $B$} = \mbox{\boldmath $R\times E$}/R.
\label{eq:LWB}
\end{equation}
Here {\boldmath $R$}
denotes the radius-vector from the retarded position
of the charge
$q$ to the point where the fields are calculated and
{\boldmath $v$} and {\boldmath $a$}
denote the retarded velocity and acceleration of $q$,
respectively. Analogous expressions hold for fields of monopoles.

    The first term on the right hand side of $(\!\!~\ref{eq:LWE})$
and of $(\!\!~\ref{eq:LWB})$ is called a velocity field and the
corresponding second term is called acceleration field
(see [7], p.657). The
acceleration fields decrease like $R^{-1}$. Hence, the
energy density of these fields $T^{00}=(E^2+B^2)/8\pi$
decreases like $R^{-2}$. The same property holds for all entries
of the energy-momentum tensor $(\!\!~\ref{eq:EMT})$
of acceleration fields. It follows that these fields represent
an objective physical entity that is endowed with an
outgoing flux of energy and momentum. On the other hand,
energy density of velocity fields as well as the
interaction energy density of velocity fields and
acceleration fields decrease like $R^{-4}$ and $R^{-3}$,
respectively. It follows that the energy flux of these
fields vanishes at infinity. This discussion shows that
acceleration fields are dominant at
very far regions and represent
free electromagnetic waves. In quantum mechanics they take
the form of real photons. The portion of space which is very far
from the origin is the wave zone.
Henceforth, acceleration fields at the wave zone are
called free waves.

    Let us examine free waves.
Consider a monochromatic plane electromagnetic wave
emitted from a very far charge (see [8] pp. 114,115). The wave
moves in the z-direction and its fields are
\begin{equation}
\mbox{\boldmath $E$} = (1,0,0)sin[\omega (t-z)]
\label{eq:EWAVE}
\end{equation}
\begin{equation}
\mbox{\boldmath $B$} = (0,1,0)sin[\omega (t-z)].
\label{eq:BWAVE}
\end{equation}
Relation $(\!\!~\ref{eq:POTE})$ shows that
these fields, which are written here in a 3-vector notation,
can be derived from the following 4-potential
\begin{equation}
A_{(e)\mu } = (0,1,0,0)cos[\omega (t-z)]/\omega .
\label{eq:AWAVE}
\end{equation}
As seen from the components of {\boldmath $E$} and of
the 4-potential $(\!\!~\ref{eq:AWAVE})$, this field
is polarized in the x-direction.

    Let us examine how the free wave whose fields are
$(\!\!~\ref{eq:EWAVE})$ and $(\!\!~\ref{eq:BWAVE})$
is seen in the $\{m\}$ system. For this purpose
consider the 4-potential
\begin{equation}
A_{(m)\mu } = (0,0,1,0)cos[\omega (t-z)]/\omega .
\label{eq:MAWAVE}
\end{equation}
An application of $(\!\!~\ref{eq:POTM})$ and of the notation
$(\!\!~\ref{eq:FMMUNU})$ of the entries of the monopole
fields tensor, shows that the electric and the magnetic
fields $(\!\!~\ref{eq:EWAVE})$ and $(\!\!~\ref{eq:BWAVE})$
are obtained from the 4-potential $(\!\!~\ref{eq:MAWAVE})$.
On the basis of these results it is
concluded here that free waves of charges
look the same as free waves of monopoles.

    Let us turn to velocity fields of the systems. The problem
of interaction of monopoles with velocity fields of charges
is analyzed by means of the following example. Consider
two magnetic polar dipoles made of monopole matter. One face
of each dipole is covered with $+g$ magnetic monopoles and
the opposite face is covered with $-g$ magnetic monopoles.
Let $p$ denote the strength of the dipoles.
In a coordinate frame $\Sigma $,
the dipoles are placed as depicted in Fig. 1.a. The force
exerted on the upper dipole is the same as the force
found in an analogous system of polar electric dipoles[9].
Using also the well known expression for dipole field
(see [7], p. 141), one finds the required force
\begin{equation}
\mbox{\boldmath $f$} = (0,-1,0)3p^2/(8r^4)
\label{eq:FDIP}
\end{equation}
The force exerted on the lower dipole has the same size
and takes the opposite direction. It follows that the
two dipoles attract each other.

    Let us examine the same system in another coordinate
frame $\Sigma '$.  This frame is related to $\Sigma $ by
means of space reflection. Thus, $\Sigma $ is a right handed
frame and $\Sigma '$ is a left handed frame. Obviously, the
frame $\Sigma '$, like $\Sigma $, is a legitimate frame and
the laws of physics should hold in it.

    On the basis of the tensorial relations
$(\!\!~\ref{eq:LORM})$-$(\!\!~\ref{eq:POTM})$
it is found that, in
$\Sigma '$, the dipoles are seen as depicted in
fig. 1.b. Using the law of force for the dipoles of
fig. 1.b, one finds that they attract each other with a
force $(\!\!~\ref{eq:FDIP})$. This outcome is an example
of the consistency of the laws of physics as seen in the
two coordinate frames $\Sigma $ and $\Sigma '$.

    Let us return to the frame $\Sigma $ and examine two
axial magnetic dipoles made of current loops (see fig. 2.a).
Each of these dipoles replaces one of the dipoles of
fig. 1.a and the strength of these dipoles satisfies
$M=p$.  It turns out that the force exerted on each axial dipole
is the same as the force found for the polar dipoles of
the previous example (see [7], p. 185). It follows that,
like in the case of the polar dipoles depicted in fig. 1.a, the
two axial dipoles of fig. 1.b attract each other.

    In the second frame $\Sigma '$, the current loops are
seen as depicted in fig. 2.b. Using the law of force, it is
easily seen that in $\Sigma '$, like in $\Sigma $, the two
dipoles attract each other. Thus, the outcome of the
analysis of the two axial dipoles is another example
of the consistency of physics.

    Let us turn to a hybrid system made of one polar magnetic
dipole and one axial magnetic dipole. In the frame $\Sigma $
these dipoles are seen as in fig. 3.a. In this case we examine
the interaction of monopoles with velocity fields of charges
and vice versa. Let us test the assumption stating:
\begin{itemize}
\item[{C.}] Monopoles interact with velocity fields of charges
in the same way as they do with velocity fields of monopoles.
\end{itemize}
Hereafter, this assumption is called assumption C.

    Using assumption C, one finds that the polar dipole, which
is made of two displaced monopoles $\pm g$, is seen in
$\Sigma $ at {\boldmath $r$}$_1 =(0,0,r)$.
This dipole is pooled downwards,
like its counterpart of fig. 1.a. Requiring momentum conservation,
one finds that the axial dipole seen in fig. 3.a at
{\boldmath $r$}$_2 =(0,0,-r)$ should be pooled upwards, like
its counterpart
of fig. 2.a. Thus, it is seen that in $\Sigma $ the two dipoles of
fig. 3.a {\em attract} each other.

    Let us turn to the frame $\Sigma '$. In this frame the
polar dipole is seen in {\boldmath $r$}$'_1 =(0,0,-r)$.
Using the tensorial relations
$(\!\!~\ref{eq:LORM})$-$(\!\!~\ref{eq:POTM})$,
one finds that
its dipole moment points downwards. This is the same
situation as seen in fig. 1.b. The axial dipole, which is
made of a current loop, is seen in $\Sigma '$ at
{\boldmath $r$}$'_2 =(0,0,r)$ and its dipole moment
points upwards.  Thus, it is shown that in
$\Sigma '$, the directions of the dipoles are anti-parallel
whereas they are parallel in $\Sigma $.
Using assumption C, one finds that, in $\Sigma '$,
the two dipoles should {\em repel} each other. This result
contradicts the findings of the right handed frame $\Sigma $.

    The contradiction obtained above disproves assumption C.
The problem can be settled if one concludes:
\begin{itemize}
\item[{D.}] Monopoles do not interact with velocity
fields of charges and charges do not interact with velocity fields
of monopoles. Charges and monopoles interact with
free waves of both systems.
\end{itemize}

    Conclusion D is consistent with the results of the
experiment described in fig. 3. Indeed, in the situation
shown in fig. 3.a, as well as in that of fig. 3.b, one finds
that if conclusion D holds then no force is exerted on the
dipoles. Thus, the frame $\Sigma $ and $\Sigma '$ yield
compatible results.

    Conclusion D is consistent with the charge-monopole equations
of motion obtained in [4] under assumptions A and B, as
presented in the introduction. Thus, the particles, equation
of motion are
\begin{equation}
\frac {dP^\mu _{(e)}} {d\tau } = F_{(e,w)}^{\mu \nu }J_{(e)\nu },
\label{eq:LOREF}
\end{equation}
and
\begin{equation}
\frac {dP^\mu _{(m)}} {d\tau } = F_{(m,w)}^{\mu \nu }J_{(m)\nu },
\label{eq:LORMF}
\end{equation}
Here $F_{(e,w)}^{\mu \nu }$ represents the velocity fields
of charges, their acceleration fields and free waves emitted
from monopoles.
Its entries are like those of $(\!\!~\ref{eq:FMUNU})$.
$F_{(m,w)}^{\mu \nu }$ is defined analogously.
Its entries are like those
of $(\!\!~\ref{eq:FSMUNU})$ and $(\!\!~\ref{eq:FMMUNU})$.

    It is found above that only free waves are
common to charges and monopoles. These fields satisfy
homogeneous Maxwell equations. Thus, Maxwell equations
$(\!\!~\ref{eq:MAXEIH})$, $(\!\!~\ref{eq:MAXEH})$,
$(\!\!~\ref{eq:MAXMIH})$ and $(\!\!~\ref{eq:MAXMH})$
are not affected.

\vglue 0.6666666in
\noindent
{\bf 4. Discussion}
\vglue 0.33333in

    The equations of motion, namely, the Lorentz force
$(\!\!~\ref{eq:LOREF})$ and $(\!\!~\ref{eq:LORMF})$
as well as Maxwell equations
$(\!\!~\ref{eq:MAXEIH})$, $(\!\!~\ref{eq:MAXEH})$,
$(\!\!~\ref{eq:MAXMIH})$ and $(\!\!~\ref{eq:MAXMH})$
are the same as those obtained
in [4] by means of assumption B which states that the
theory should be derived from a regular Lagrangian function.
Two conclusions can be deduced from this outcome:
\begin{itemize}
\item[{I. }] Requirement B is not essential because the same results
are obtained without its utilization.
\item[{II.}] The equations of motion
can be derived from a regular Lagrangian function.
\end{itemize}
Conclusion II is obvious. Indeed, it is shown in [4] that the
Lagrangian function (written here in the metric of the present work)
\begin{eqnarray}
L =
-\frac {1}{16\pi }F_{(e,w)}^{\mu \nu }
F_{(e,w)\mu \nu }
-J_{(e)}^\mu A_{(e,w)\mu }
-\frac {1}{16\pi }F_{(m,w)}^{\mu \nu }
F_{(m,w)\mu \nu }
\nonumber \\
-J_{(m)}^\mu A_{(m,w)\mu }
+\frac {1}{16\pi }(F_{(w)}^{\mu \nu }
F_{(w)\mu \nu }
\label{eq:LAGRAN}
\end{eqnarray}
yields the equations of motion obtained here.

    Another result of the analysis carried out in the previous
section is that an alternative
charge-monopole theory whose equations
of motion differ from those derived above,
must be incompatible with at least one of the postulates
used here. Consider, for example, a theory that does
not distinguish between electromagnetic fields of
charges and those of monopoles and uses the
following particles' equations of motion[1]
\begin{equation}
\frac {dP^\mu _{(e)}} {d\tau } = F_{(e,m,w)}^{\mu \nu }J_{(e)\nu },
\label{eq:DIRACE}
\end{equation}
and
\begin{equation}
\frac {dP^\mu _{(m)}} {d\tau } = F_{(e,m,w)}^{*\mu \nu }J_{(m)\nu }.
\label{eq:DIRACM}
\end{equation}

    It turns out that in this theory a pure monopole
system does not have the ordinary structure of
electrodynamics. Here monopole fields are not defined like in
$(\!\!~\ref{eq:LORM})$ as done in electrodynamics
but are postulated to have the same
physical meaning as fields of
charges and are obtained from the later by means
of the duality transformation $(\!\!~\ref{eq:FSMUNU})$.
As a result, this theory cannot use potentials
of the form $(\!\!~\ref{eq:POTM})$ that yield a tensor
whose entries are like $(\!\!~\ref{eq:FMMUNU})$. Moreover,
{\em regular} potentials of electrodynamics of charges
$(\!\!~\ref{eq:POTE})$ cannot be used[10]
for monopole fields because,
as is well known, in electrodynamics of charges, where fields are
derived from potentials like $(\!\!~\ref{eq:POTE})$
\begin{equation}
\mbox{\boldmath $\nabla \cdot B$} =
\mbox{\boldmath $\nabla \cdot $}(\mbox{\boldmath $\nabla \times A$}) = 0.
\label{eq:DIVB}
\end{equation}
This property of {\em regular} potentials is inconsistent with
\begin{equation}
\mbox{\boldmath $\nabla \cdot B$} = 4\pi \rho_{(m)}
\label{eq:DIVBMON}
\end{equation}
which is the $\mu=0$ equation of $(\!\!~\ref{eq:MAXMIH})$.

    As is well known, potentials are essential ingredients of
the electrodynamic Lagrangian function. It follows that the
equations of motion $(\!\!~\ref{eq:DIRACE})$
and $(\!\!~\ref{eq:DIRACM})$ cannot be derived from a regular
Lagrangian. This conclusion holds also in the limit
of a system of monopoles which does not include electric charges.

    Another aspect of the theory that uses $(\!\!~\ref{eq:DIRACE})$
and $(\!\!~\ref{eq:DIRACM})$  as the equations of motion of
charges and monopoles, respectively, is the assignment of
pseudo-scalar properties to monopoles
(see [7], p. 253). Indeed, if monopoles
are scalars, as used in $(\!\!~\ref{eq:LORM})$, then
the problem presented in fig. 3 persists. On the other hand,
if monopoles are pseudo-scalars then the polar dipole of fig. 3.b,
which is made of two displaced monopoles, is not as
shown in the figure but reverses its
direction. Under these circumstances the dilemma
apparently disappears.

   Consequences of the assignment of pseudo-scalar properties
to monopoles affect the construction of a Lagrangian for the
equations of motion $(\!\!~\ref{eq:DIRACE})$
and $(\!\!~\ref{eq:DIRACM})$. Thus, here $j_{(m)}^\mu $ is
{\em not} a 4-vector.
It follows that it cannot be used for constructing
a scalar with a 4-potential $A_{(e)}^\mu $ of charge which
transforms like a 4-vector.

\vglue 0.6666666in
\noindent
{\bf 5. Concluding Remarks}
\vglue 0.33333in

    The discussion carried out in this work relies on the
postulate stating that a classical theory of a system
which consists of one kind of charges and whose equations
of motion are like those of classical electrodynamics, takes
a form which is analogous to this theory. Thus, equations
$(\!\!~\ref{eq:LORM})$-$(\!\!~\ref{eq:POTM})$ are elements
of the theory. In addition to this, it is assumed that a
theory of a combined system which consists of several kinds
of charges, reduces to the theory of one kind of charge as
formulated in $(\!\!~\ref{eq:LORM})$-$(\!\!~\ref{eq:POTM})$,
in an appropriate limiting process where
all other kinds of charges are removed to infinity.

    It is proved that these assumptions lead to the equations
of motion of a charge-monopole system
$(\!\!~\ref{eq:LOREF})$, $(\!\!~\ref{eq:LORMF})$ and to
the obvious form of Maxwell equations
$(\!\!~\ref{eq:MAXEIH})$, $(\!\!~\ref{eq:MAXEH})$,
$(\!\!~\ref{eq:MAXMIH})$ and $(\!\!~\ref{eq:MAXMH})$.
These equations
state that charges do not interact with velocity fields of
monopoles and monopoles do not interact with velocity fields
of charges. Charges and monopoles interact indirectly by
means of free electromagnetic waves (namely, real photons).

   The same equations of motion have been obtained earlier,
using a postulate stating that the theory should be derivable
from a regular Lagrangian function[4]. It is shown here that
this requirement is not essential and that the same results
are obtained without relying on the existence of a
regular Lagrangian.

    It is also proved here that any alternative theory whose
equations of motion differ from
$(\!\!~\ref{eq:LOREF})$-$(\!\!~\ref{eq:LORMF})$ cannot
satisfy the postulates used here. The theory
whose equations of motion are
$(\!\!~\ref{eq:DIRACE})$ and $(\!\!~\ref{eq:DIRACM})$,
where charges and monopoles interact with all kinds of fields,
is a particular case. Here one
uses monopoles that are pseudo-scalars
and the pure monopole part of this theory does not take the ordinary
form of electrodynamics
$(\!\!~\ref{eq:LORM})$-$(\!\!~\ref{eq:POTM})$.

\newpage
References:
\begin{itemize}
\item[{[1]}] P. Goddard and D. I. Olive, Rep. Prog. Phys.,
{\bf 41}, 1357 (1978) and references therein.
\item[{[2]}] J. R. Ficenec and V. L. Teplitz in Electromagnetism
(D. Teplitz, Editor) (Plenum, New York, 1982) and references therein.
\item[{[3]}] F. Rohrlich, Phys. Rev. {\bf 150}, 1104 (1966).
\item[{[4]}]  E. Comay, Nuovo Cimento, {\bf {80B}}, 159 (1984).
\item[{[5]}] H. J. Lipkin and M. Peshkin,
Phys. Lett. {\bf B179}, 109 (1986).
\item[{[6]}] E. Comay, Phys. Lett. {\bf B187}, 111 (1987).
\item[{[7]}] J. D. Jackson, Classical Electrodynamics (John
Wiley, New York, 1975). p. 28.
\item[{[8]}] L. D. Landau and E. M. Lifshitz, The Classical
Theory of Fields (Pergamon, Oxford, 1975). p.61.
\item[{[9]}] W. K. H. Panofsky and M. Phillips, Classical
Electricity and Magnetism (Addison Wesley, Reading Mass., 1962).
p. 19.
\item[{[10]}]  P. A. M. Dirac, Phys. Rev. {\bf {74}}, 817 (1948).
\end{itemize}
\newpage
\noindent
Figure Captions

\noindent
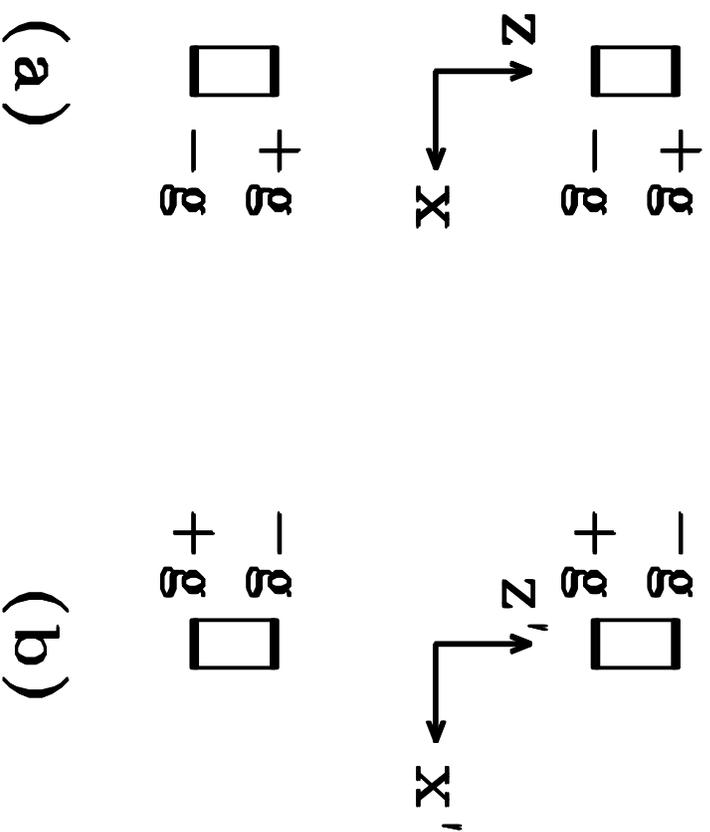
Fig. 1:

\begin{itemize}
\item[{(a)}] Two polar magnetic dipoles are held
fixed at ${\bf r}_i=(0,0,\pm r)$, respectively. The signs of
the surface monopoles $\pm g$ indicate the direction of the dipole
moment.
\item[{(b)}] The system as seen in the left handed laboratory $\Sigma '$.
\end{itemize}

\noindent
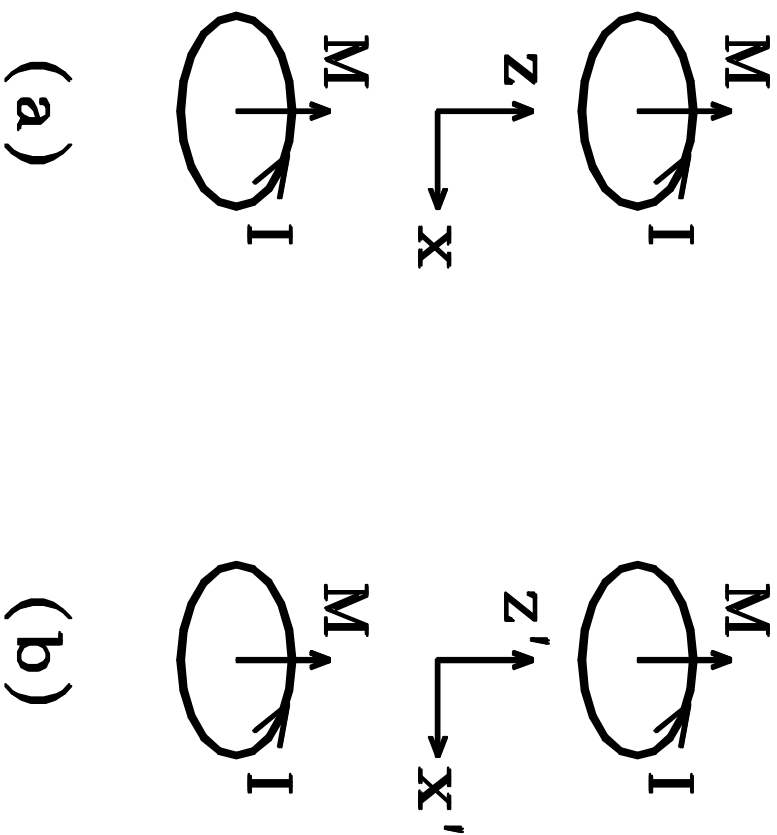
Fig. 2:

\begin{itemize}
\item[{(a)}] Two axial magnetic dipoles ${\bf M}$, made of current
carrying loops, are fixed like in fig. 1.
\item[{(b)}] The system as seen in the left handed laboratory $\Sigma '$.
\end{itemize}

\noindent
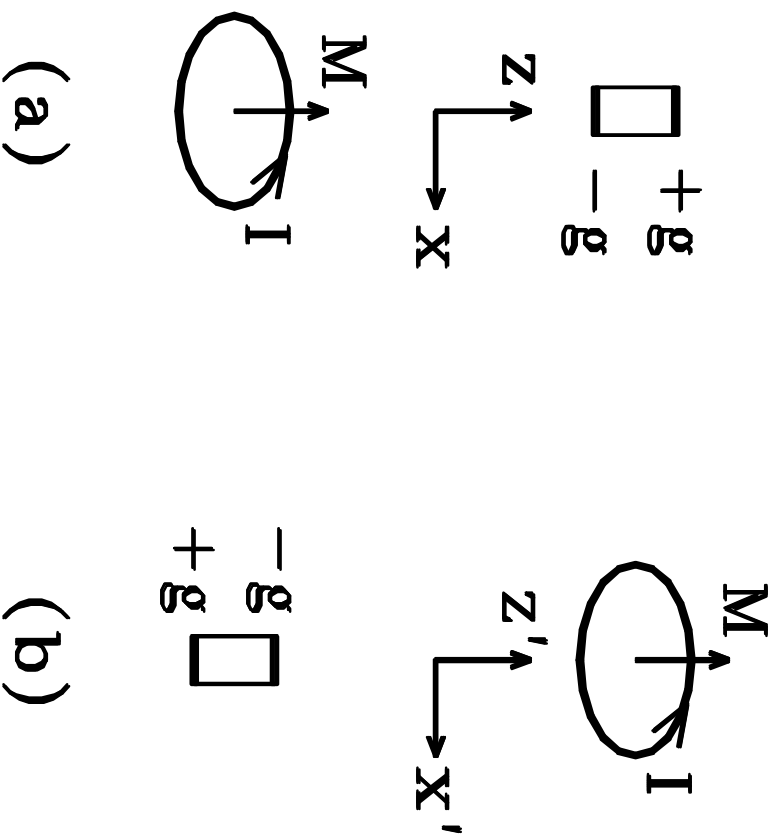
Fig. 3:

\begin{itemize}
\item[{(a)}] A polar magnetic dipole and an
axial one are fixed as shown.
\item[{(b)}] The system as seen in the left handed laboratory $\Sigma '$.
\end{itemize}

\end{document}